\documentclass[aps,prd,superscriptaddress,notitlepage]{revtex4-1}
\usepackage{amsmath}
\usepackage{bm}
\usepackage{times}
\usepackage{braket}
\usepackage{color,graphicx}
\usepackage[T1]{fontenc}
\usepackage[utf8]{inputenc}
\usepackage{hyperref}

\definecolor{nicered}{rgb}{0.7,0.1,0.1}
\definecolor{nicegreen}{rgb}{0.1,0.5,0.1}

\newcommand {\E}[1]{\times 10^{#1}}	% exponent notation
\newcommand {\e}[1]{\mathrm{\,#1}}       % units
\newcommand{\mc}[1]{\mathcal{#1}}

\newcommand{\mrm}[1]{\mathrm{#1}}

\newcommand{\bea}{\begin{eqnarray}}
\newcommand{\eea}{\end{eqnarray}}

\newcommand{\nn}[0]{\nonumber}

\newcommand{\br}[0]{\mrm{Br}}

\hypersetup{colorlinks,citecolor= nicegreen, linkcolor= nicered}

\begin{document}
\title{Is symmetry breaking of $SU(5)$ theory responsible for the diphoton excess?}

\author{Ilja Dor\v sner} \email[Electronic address:]{dorsner@fesb.hr}
\affiliation{University of Split, Faculty of Electrical Engineering, Mechanical Engineering and Naval Architecture in Split (FESB), Ru\dj era Bo\v skovi\' ca 32, 21 000 Split, Croatia}

\author{Svjetlana Fajfer} \email[Electronic
address:]{svjetlana.fajfer@ijs.si} 
\affiliation{Department of Physics,
  University of Ljubljana, Jadranska 19, 1000 Ljubljana, Slovenia}
\affiliation{J.\ Stefan Institute, Jamova 39, P.\ O.\ Box 3000, 1001
  Ljubljana, Slovenia}

\author{Nejc Ko\v snik} 
\email[Electronic address:]{nejc.kosnik@ijs.si}
\affiliation{J.\ Stefan Institute, Jamova 39, P.\ O.\ Box 3000, 1001
  Ljubljana, Slovenia}
\affiliation{Department of Physics,
  University of Ljubljana, Jadranska 19, 1000 Ljubljana, Slovenia}
    
%%%%%%%%%%%%%%%%%%%%%%%%%%%%%%%%%%%%%%%%%%%%%%%%%%
\begin{abstract}
  We advocate the possibility that the observed diphoton excess at
  $750$\,GeV at the LHC can be addressed by the scalar field that is a
  part of the $SU(5)$ symmetry breaking sector. The field in question
  is the Standard Model singlet that resides in the adjoint
  representation that breaks $SU(5)$ down to
  $SU(3) \times SU(2) \times U(1)$. We also show that the required
  production and subsequent decay to two photons of this singlet can
  be induced by individual or combined contribution of two scalar
  multiplets $S_3$ and $R_2$ that transform as
  $(\mathbf{3},\mathbf{3},-1/3)$ and $(\mathbf{3},\mathbf{2},7/6)$
  under $SU(3) \times SU(2) \times U(1)$, respectively. The individual
  dominance of these multiplets is directly related to the issue of
  the charged fermion mass generation within the $SU(5)$ framework and
  can be unambiguously tested through the diboson decay signatures of
  the Standard Model singlet field.
\end{abstract}
%%%%%%%%%%%%%%%%%%%%%%%%%%%%%%%%%%%%%%%%%%%%%%%%%%
\pacs{}
\maketitle

%%%%%%%%%%%%%%%%%%%%%%%%%%%%%%%%%%%%%%%%%%%%%%%%%%
The first results from Run 2 of the LHC experiments have revealed a hint of an unexpected feature in diphoton final state.
With integrated luminosity of $\sim 3\e{fb}^{-1}$, collected at the
center-of-mass energy of $13\e{TeV}$, both ATLAS and CMS experiments
have reported modest excesses of two-photon events over the Standard
Model~(SM) background~\cite{Aaboud:2016tru,Khachatryan:2016hje}.
%\cite{CMS-PAS-EXO-15-004,%ATLAS-CONF-2015-081}. 
The global statistical significances are small. However, local
significances of the excesses reach $3.9\,\sigma$ and $3.4\,\sigma$ at
ATLAS and CMS, respectively. These excesses are furthermore located in the same region of the diphoton invariant mass at $m_{\gamma\gamma} \simeq
750\e{GeV}$. The simplest theoretical interpretation of the
preliminary diphoton signal is to introduce a scalar
particle that is a singlet of the SM and along with it
additional fermionic and/or bosonic degrees of freedom that mediate the singlet interaction to pairs of gauge bosons. See Refs.~\cite{Knapen:2015dap,Buttazzo:2015txu,Franceschini:2015kwy} for
explicit examples.

We advocate the possibility that the SM singlet in question is a part of the $SU(5)$ symmetry breaking sector. Recall, $SU(5)$ is broken down to $SU(3) \times SU(2) \times U(1)$ through a vacuum expectation value (VEV) of the SM singlet field in $24$-dimensional scalar representation~\cite{Georgi:1974sy}. The decomposition of the adjoint representation of $SU(5)$ under $SU(3) \times SU(2) \times U(1)$ is $\mathbf{24} \equiv \Sigma = (\mathbf{8},\mathbf{1},0) \oplus (\mathbf{1},\mathbf{3},0) \oplus (\mathbf{3},\mathbf{2},-5/6) \oplus (\overline{\mathbf{3}},\mathbf{2},5/6) \oplus (\mathbf{1},\mathbf{1},0) = (\Sigma_8, \Sigma_3, \Sigma_{3,2},\Sigma_{\overline{3},2},\Sigma_0)$. Our first goal is to demonstrate that the scalar singlet $\Sigma_0 \equiv (\mathbf{1},\mathbf{1},0)$ can reside at the electroweak scale if needed. 

The scalar potential $V$ for $\Sigma$ is
\begin{equation}
\label{potential} 
V = - \frac{\mu^2}{2}
\Sigma^i_{~j} \Sigma^j_{~i} +\frac{a}{4} (\Sigma^i_{~j}
\Sigma^j_{~i})^2+\frac{b}{2} \Sigma^i_{~j}
\Sigma^j_{~k}\Sigma^k_{~l} \Sigma^l_{~i}+\frac{c}{3}
\Sigma^i_{~j} \Sigma^j_{~n} \Sigma^n_{~i},
\end{equation}
where $\mu^2$, $a$, $b$, and $c$ represent parameters of the
theory. $i,j,k,l,n=1,\ldots,5$ are the $SU(5)$ indices. We, for
definiteness, consider only renormalizable operators. The conditions
that the potential $V$ develops a local minimum that
breaks the $SU(5)$ down to the SM gauge group are~\cite{Guth:1981uk}
\begin{equation}
\beta > \left\{
\begin{array}{rc}
\frac{15}{32}\left(\gamma-\frac{4}{15}\right), & \gamma>\frac{2}{15} \\
\\
-\frac{1}{120 \gamma}, & \gamma<\frac{2}{15} \\
\end{array} \right.,
\end{equation}
where dimensionless variables $\beta$ and $\gamma$ are defined as $
\beta=(\mu^2 b)/c^2$ and $\gamma=(a/b+7/15)$, respectively. The symmetry breaking VEV of $\Sigma$ is $\langle
\Sigma\rangle=\lambda/\sqrt{30} \ \textrm{diag}(2,2,2,-3,-3)$,
where~\cite{Guth:1981uk}
\begin{equation}
\lambda=\frac{c}{b} \left(\frac{\beta}{\gamma}
\right)^{1/2} \left[\left(1+\frac{1}{120 \beta
\gamma}\right)^{1/2}+\frac{1}{(120 \beta
\gamma)^{1/2}}\right]=\frac{c}{b}
\left(\frac{\beta}{\gamma} \right)^{1/2} h(\beta \gamma).
\end{equation}

$\Sigma_{3,2}$ and $\Sigma_{\overline{3},2}$ multiplets are eaten by $X$ and $Y$ gauge bosons of $SU(5)$. These gauge fields mediate proton decay and thus need to be very heavy. Their common mass $m_{(X,Y)}$ is 
\begin{equation}
m_{(X,Y)}=\sqrt{\frac{5}{12}} \ g_\mathrm{GUT} \lambda,
\end{equation}
where $g_\mathrm{GUT}$ is the $SU(5)$ gauge coupling at the grand
unified theory (GUT) scale $m_\mathrm{GUT}$. It is customary to
identify $m_{(X,Y)}$ to be the GUT scale, i.e., scale where the SM
gauge couplings unify. Potential in Eq.~\eqref{potential} yields the
following mass relations
\begin{equation}
  \begin{split}
%\begin{eqnarray}
m^2_{\Sigma_8}&= \left[\frac{1}{3}+\frac{5}{\sqrt{30}}
\left(\frac{\gamma}{\beta} \right)^{1/2} \frac{1}{h(\beta \gamma)}
\right] b \lambda^2,\\
m^2_{\Sigma_3}&= \left[\frac{4}{3}-\frac{5}{\sqrt{30}}
\left(\frac{\gamma}{\beta} \right)^{1/2} \frac{1}{h(\beta \gamma)}
\right] b \lambda^2,\\
m^2_{\Sigma_{0}}&= \left[1-\frac{1}{1+(1+120 \beta
\gamma)^{1/2}}\right] 2 b \gamma \lambda^2,
%\end{eqnarray}
  \end{split}
\end{equation}
where $m_{\Sigma_8}$, $m_{\Sigma_3}$, and $m_{\Sigma_{0}}$ denote masses of $\Sigma_8$, $\Sigma_3$, and $\Sigma_{0}$, respectively.

We require $\Sigma_{0}$ to be light. We accordingly set $m^2_{\Sigma_{0}} = 2 b \gamma \lambda^2 \epsilon$ to demonstrate viability of this requirement, where $0<\epsilon \ll 1$. This leads to the following inequality (for $\gamma>0$) 
\begin{equation}
\beta \approx \frac{\epsilon^2-1}{120 \gamma} > -\frac{1}{120 \gamma}.
\end{equation}
We furthermore obtain $m^2_{\Sigma_8}= [1/3+10 \gamma - \mathcal{O}(\epsilon)] b \lambda^2$ and $m^2_{\Sigma_3}= [4/3- 10 \gamma
+ \mathcal{O}(\epsilon)] b \lambda^2$. Clearly, the requirement that $m^2_{\Sigma_3}>0$ is satisfied for $\gamma<2/15$. This shows that there exists a part of the parameter space where $\Sigma_{0}$ can reside at the electroweak scale to serve as the candidate behind the diphoton excesses. This possibility is not in collision with the symmetry breaking chain $SU(5)\rightarrow SU(3) \times SU(2) \times U(1)$ at the classical level.

Singlet field $\Sigma_0$ of mass $m_{\Sigma_0} \simeq 750$\,GeV should couple to vector-like fermions and/or charged scalars in order to be produced at the LHC and to be able to subsequently decay into two
photons. Only then will it be able to help explain observed signal
excesses~\cite{Knapen:2015dap,Buttazzo:2015txu,Franceschini:2015kwy}. Vector-like quarks and leptons are frequently used
in GUT model building to address, for example, the issue of the SM
fermion masses and mixings. The idea is to mix the SM fermions with
one or more of the SM multiplets in these additional $SU(5)$
representations to produce viable masses and mixing parameters~\cite{Witten:1979nr}. The most commonly used
representations to accommodate vector-like states are $5$-, $10$-,
$15$-, and $24$-dimensional representations. The relevant operators,
at the $SU(5)$ level, are straightforward to write down and we omit them in this note. For explicit proposals to couple vector-like representations to an $SU(5)$ singlet to address observed diphoton excess at $750$\,GeV at the LHC in a non-supersymmetric (supersymmetric) setting see Ref.~\cite{Patel:2015ulo} (Refs.~\cite{Hall:2015xds,Dutta:2016jqn}). The use of vector-like multiplets that comprise full $SU(5)$ multiplet(s) of $5$- and $10$-dimensional nature has also been advocated in Ref.~\cite{Ellis:2015oso}.  

Scalar states, on the other hand, are necessary since one or more Higgs doublets are needed to generate fermion masses in the first place. Most commonly used scalar representations in $SU(5)$ are accordingly $5$- and $45$-dimensional ones. The latter representation contains, among other states, two scalar multiplets $S_3$ and $R_2$ that transform as $(\mathbf{3},\mathbf{3},-1/3)$ and $(\mathbf{3},\mathbf{2},7/6)$ under $SU(3) \times SU(2) \times U(1)$, respectively. (Here we adopt notation of Ref.~\cite{Buchmuller:1986zs} to denote relevant colored scalar multiplets.) These particular fields can generate required signal strength very efficiently as we show next. 

The operators that couple $S_3$ and $R_2$ to the singlet $\Sigma_0$, at the $SU(5)$ level, are $m \mathbf{45}^{i j}_k \Sigma^k_{~l} \mathbf{45}_{i j}^{l\,*}$ and $\sigma \mathbf{45}^{i j}_k \Sigma^k_{~l} \Sigma^l_{~n} \mathbf{45}_{i j}^{n\,*}$, where $m$ and $\sigma$ are {\it a priori} unknown dimensionful and dimensionless coefficients, respectively. We find that the relevant trilinear vertex is $(-m\sqrt{3/10}+3/5 \sigma \lambda) \Sigma_0 (S_3^\dagger S_3+R_2^\dagger R_2)$. The trilinear vertex coefficient is thus the same for both $S_3$ and $R_2$ due to the underlying $SU(5)$ symmetry. 

From the point of view of an effective theory defined at the $1$\,TeV scale the most important operator for our phenomenological study is
\begin{equation}
\label{eq:PhenoLag}
  \mathcal{L}^{\rm eff} \supset x m_{\Sigma_0} \Sigma_0
  \left(S_3^\dagger S_3 + R_2^\dagger R_2\right), \end{equation}
where we set $m_{\Sigma_0} = 750\e{GeV}$. As shown in the previous paragraph, dimensionless parameter $x$ is directly related to the parameters of the GUT potential. The trilinear vertex
of Eq.~\eqref{eq:PhenoLag} destabilizes $\Sigma_0$ by opening a decay channel to a
pair of $S_3$'s and/or $R_2$'s, if these are lighter than
$m_{\Sigma_0}/2$. We, however, opt to present our analysis in the regime
where $\Sigma_0$ cannot decay to a pair of on-shell colored scalar states.
In such a setting $\Sigma_0$ decays predominantly to the pairs of the
SM gauge bosons via loops containing electrically charged colored scalars. 

We adapt the analogous expressions for the SM Higgs decay
widths for $h \to \gamma\gamma, gg$, in the presence of scalar degrees of freedom,
to a particular case of $\Sigma_0  \to \gamma \gamma,
gg$~\cite{Djouadi:2005gj,Dorsner:2012pp}. (See also
Appendix~\ref{sec:appendix} for more details.) In the case that the $R_2$ contribution is dominant the width expressions read
\begin{equation}
  \label{eq:Widths}
\begin{split}
\Gamma(\Sigma_0 \to \gamma\gamma)& = |x|^2\frac{\alpha^2 m_{\Sigma_0}^5}{2^{10} \pi^3
  m_{R_2}^4} D^{R_2}_{\gamma\gamma}
|\mc{A}_0(\tau)|^2,\\
\Gamma(\Sigma_0 \to gg) &=|x|^2 \frac{\alpha_S^2 m_{\Sigma_0}^5}{2^{5} \pi^3
  m_{R_2}^4} C(R_2)^2|\mc{A}_0(\tau)|^2.
\end{split}
\end{equation}
We will also consider regime in which $S_3$
and $R_2$ are simultaneously affecting the $\Sigma_0$ decays and in order to do that we take into account the decay amplitudes with interference effects included. The charge eigenstates within weak multiplets $S_3$ and $R_2$ are assumed to be degenerate with
a common masses of $m_{S_3}$ and $m_{R_2}$, respectively. We denote by $\alpha$ ($\alpha_S$) the
electromagnetic (strong) coupling, the color algebra factor for color
triplets is $C(S_3) = 1/2$, whereas 
$D_{\gamma\gamma}$ represents the boost factor of the diphoton width
stemming from the sum over all charge and color eigenstates
propagating in the loop. $D_{\gamma\gamma}$ reads
\begin{equation}
  D_{\gamma\gamma} = \left\{d_c (2T+1)\left[Y^2+\frac{T(T+1)}{3}\right]\right\}^2,
\end{equation}
where $d_c$ is the dimension of the $SU(3)$ representation of the scalar, $Y$
is its hypercharge and $T$ the weak isospin. Given the strong
dependence of $D_{\gamma\gamma}$ on hypercharge and weak isospin it is
now evident why we favor at least one of the two scalar triplets with
$d_c = 3$, namely $S_3$ or $R_2$, to be light. Their SM quantum numbers
--- $Y=-1/3$, $T=1$ for $S_3$ and $Y=7/6$, $T=1/2$ for $R_2$ --- yield
large diphoton boost factors ($D^{S_3}_{\gamma\gamma} = 49$, $D^{R_2}_{\gamma\gamma} \approx 93$)
compared to majority of other scalars contained within the $45$-dimensional representation.
For a more vivid comparison, consider
colored scalar in the representation $(\overline{\mathbf{3}},\mathbf{1},1/3)$, studied in
Ref.~\cite{Bauer:2015boy}, or $(\mathbf{3},\mathbf{2},1/6)$ that result in
$0.1$ and $2.8$ for $D_{\gamma\gamma}$, respectively. The loop
function of the argument $\tau = m_{\Sigma_0}^2/(4 m_\mrm{LQ}^2)$,
with $\mrm{LQ} = S_3, R_2$, reads
\begin{equation}
\mc{A}_0(\tau) =\frac{f(\tau)-\tau}{\tau^2},\qquad f(\tau) = \left\{
  \begin{array}{ccl}
    \arcsin^2\sqrt{\tau}& ; & \tau\leq 1,\\
    -\tfrac{1}{4}\left(\log\tfrac{1+\sqrt{1-1/\tau}}{1-\sqrt{1-1/\tau}}-i
    \pi\right)^2& ; & \tau >1
  \end{array}\right. ,
\end{equation}
and is consistent with the decay amplitude expressions we present in Appendix~\ref{sec:appendix}.

The gluonic decay $\Sigma_0 \to gg$ dominates the total width
$\Gamma_{\Sigma_0}$ while the remaining diboson widths are subleading
but non-negligible. We accordingly included widths for $\Sigma_0 \to
Z\gamma, ZZ, WW$ processes in the total width $\Gamma_{\Sigma_0}$ in our
analysis. The ratios of the diboson to diphoton decay widths that we
list in Table~\ref{tab:dib}, for two cases, where either one or the
other of the two colored scalars is dominant, exhibit very little dependence on the colored state mass. Closer inspection of Table~\ref{tab:dib} reveals that one could clearly distinguish the two scenarios through the decays of $\Sigma_0$ into diboson channels. (Here and in the following we employ for the gauge
couplings $\alpha_S(m_{\Sigma_0}/2) = 0.095$~\cite{Chatrchyan:2013txa}
and $\alpha(m_Z) = 0.0078$~\cite{Agashe:2014kda}.)
\begin{table}[!htbp]
  \centering
  \begin{tabular}{|c|c|c|c|c|}\hline
    $V V'$& $Z\gamma$ & $ZZ$ &$W^+W^-$ & $gg$\\\hline\hline 
    $\tfrac{\Gamma(\Sigma_0 \to V V')}{\Gamma(\Sigma_0 \to \gamma
    \gamma)}\big|_{m_{S_3} \ll m_{R_2}}$ & $4.3$ & $7.8$ & $26$ & $54$\\\hline
    $\tfrac{\Gamma(\Sigma_0 \to V V')}{\Gamma(\Sigma_0 \to \gamma
    \gamma)}\big|_{m_{R_2} \ll m_{S_3}}$ & $0.062$ & $0.55$ & $0.85$ & $13$\\\hline
    $\tfrac{\Gamma(\Sigma_0 \to V V')}{\Gamma(\Sigma_0 \to \gamma
    \gamma)}\big|_{m_{R_2} = m_{S_3}}$ & $0.52$ & $2.6$ & $7.2$ & $27$\\\hline
  \end{tabular}
  \caption{Ratio of diboson to diphoton decay widths $\Gamma(\Sigma_0 \to V V')/\Gamma(\Sigma_0 \to \gamma
    \gamma)$. The predictions
    of the $S_3$ ($R_2$) dominance case is shown in the first (second)
    numeric row. Results in the last row are obtained with the
    assumption of the mass degeneracy for the two colored scalar states.}
  \label{tab:dib}
\end{table}

\begin{figure}[!htb]
  \centering
  \begin{tabular}{lr}
  \includegraphics[width=0.45\textwidth]{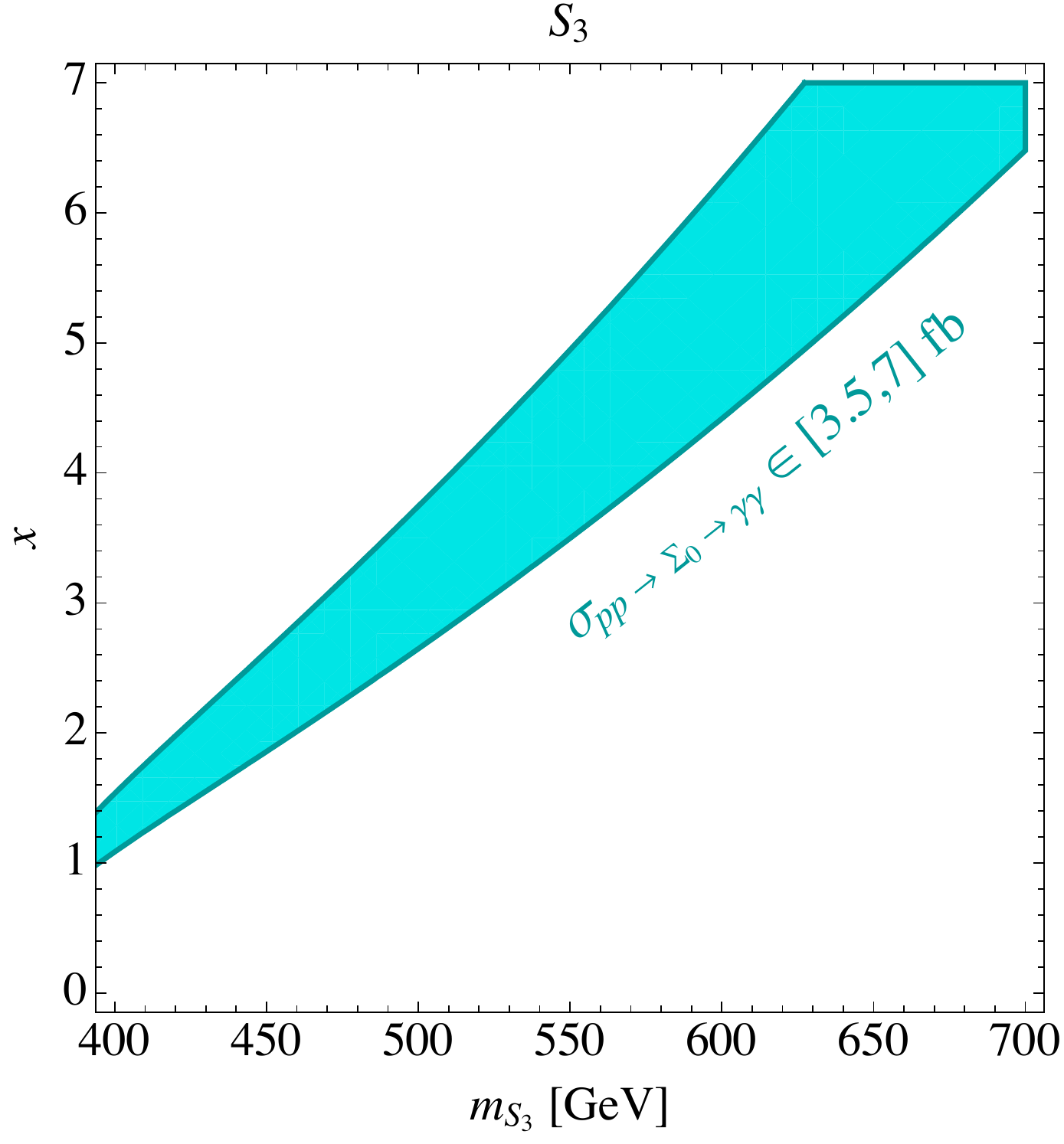}  & \includegraphics[width=0.45\textwidth]{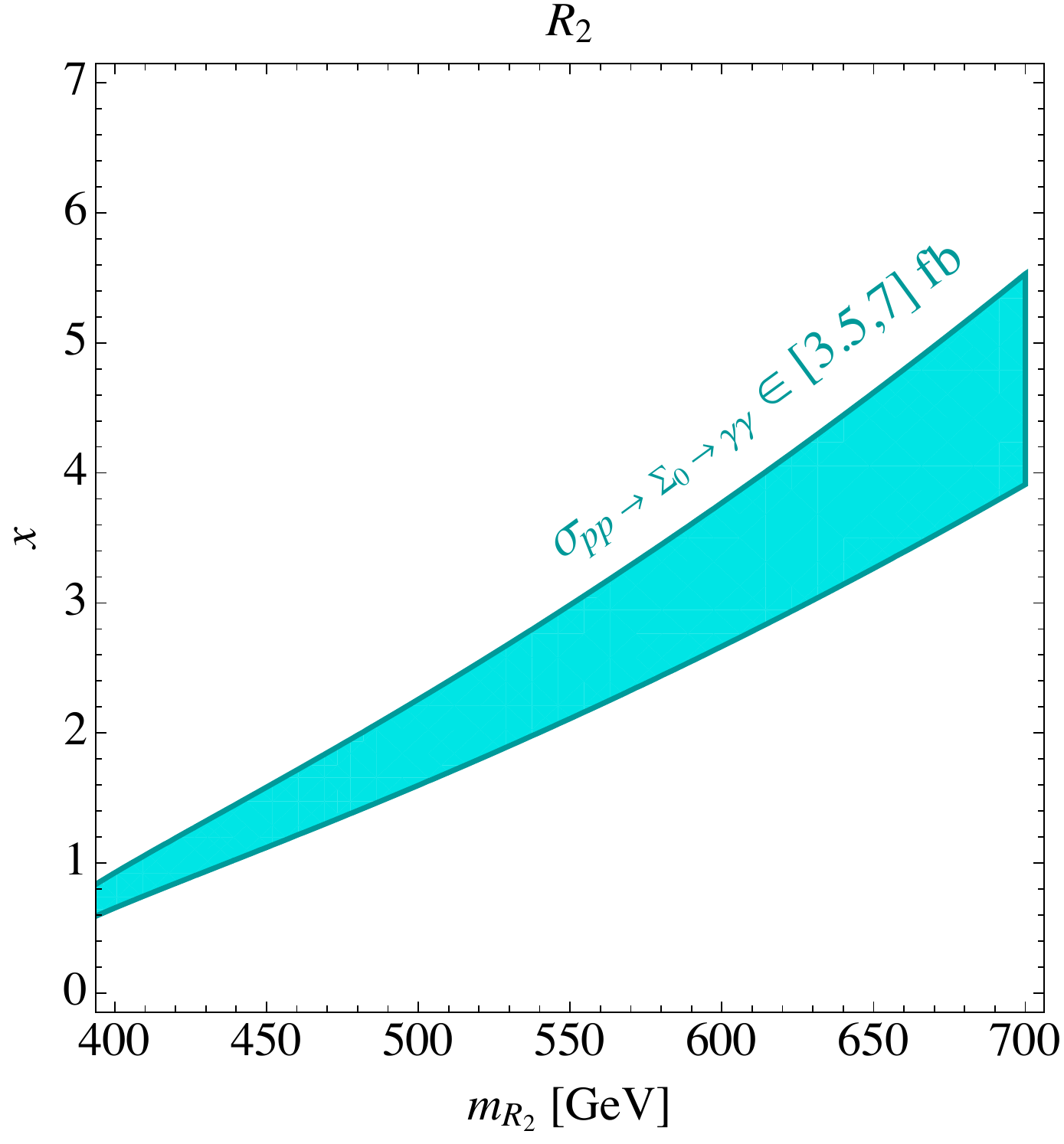}
  \end{tabular}  
  \caption{Left panel: parameter space for the case of $S_3$ dominance
    in the $m_{S_3}$--$x$ plane that satisfies the constraint on
    $\sigma (pp \to \Sigma_0 \to \gamma\gamma) \in [3.5,7]\e{fb}$ in
    cyan. Right panel: parameter space for the case of $R_2$ dominance.}
  \label{fig:S3R2}
\end{figure}

Several phenomenological analyses revealed the main characterizing feature of the excess observed in $\sigma(pp \to \gamma
\gamma)_{m_{\gamma\gamma} \approx 750\e{GeV}}$ at $\sqrt{s} =
13\e{TeV}$ (see
e.g.~\cite{Franceschini:2015kwy,Ellis:2015oso,Franceschini:2016gxv}). Assuming
a narrow scalar diphoton resonance we employ the following value in
this work:
\begin{equation}
\label{eq:Excess}
\sigma (pp \to \Sigma_0) \br (\Sigma_0 \to \gamma \gamma) \approx (3.5-7)\e{fb}.
\end{equation}
Recorded statistics in ATLAS and CMS datasets are insufficient at the
moment to be able to determine the width of the $\Sigma_0$ resonance.
Good consistency with the dataset is obtained both for large
$\Gamma_{\Sigma_0} \lesssim 0.1 m_{\Sigma_0}$, as well as for
significantly narrower $\Gamma_{\Sigma_0}$~\cite{Ellis:2015oso,ATLAS-CONF-2016-018,CMS-PAS-EXO-16-018}.
We can relate the diphoton excess of Eq.~\eqref{eq:Excess} with the partial decay
widths
\begin{equation}
  \sigma (pp \to \Sigma_0) \br (\Sigma_0 \to \gamma \gamma) =
  \frac{K_{gg} C_{gg}}{m_{\Sigma_0} s} \frac{\Gamma(\Sigma_0 \to gg)
    \Gamma(\Sigma_0 \to \gamma\gamma)}{\Gamma_{\Sigma_0}},
\end{equation}
where we employ factor $K_{gg}$ to include higher order QCD
corrections, whereas the gluon parton distribution function
convolution is embodied in $C_{gg}$. At center-of-mass energy of the
LHC Run 2 of $\sqrt{s} = 13\e{TeV}$ we adopt $C_{gg} = 2.1\E{3}$ and
$K_{gg} \approx 1.5$, where both values are taken from
Ref.~\cite{Franceschini:2015kwy}. We show in the left (right) panel of
Fig.~\ref{fig:S3R2} the region in $m_{S_3}$--$x$ ($m_{R_2}$--$x$)
plane that satisfies the constraint given in Eq.~\eqref{eq:Excess}
assuming that only state $S_3$ ($R_2$) is light. The allowed regions
presented in Fig.~\ref{fig:S3R2} suggest that relatively light colored
scalars with coupling $x$ of order $1$ can individually accommodate
observed signal. From Table~\ref{tab:dib} one can observe that other
diboson partial widths are enhanced (suppressed) with respect to the
diphoton partial width in the $S_3$ ($R_2$) dominance
scenario. Current searches at the LHC are not yet sensitive to other
diboson decays of $\Sigma_0$~\cite{Franceschini:2015kwy,
  Gupta:2015zzs} but this particular feature might be experimentally
accessible in near future. This would allow one to probe the nature of
the source of the diphoton excess.

%%%%%% %%%%%%%%% 

\begin{figure}[!htb]
  \centering
  \includegraphics[width=0.5\textwidth]{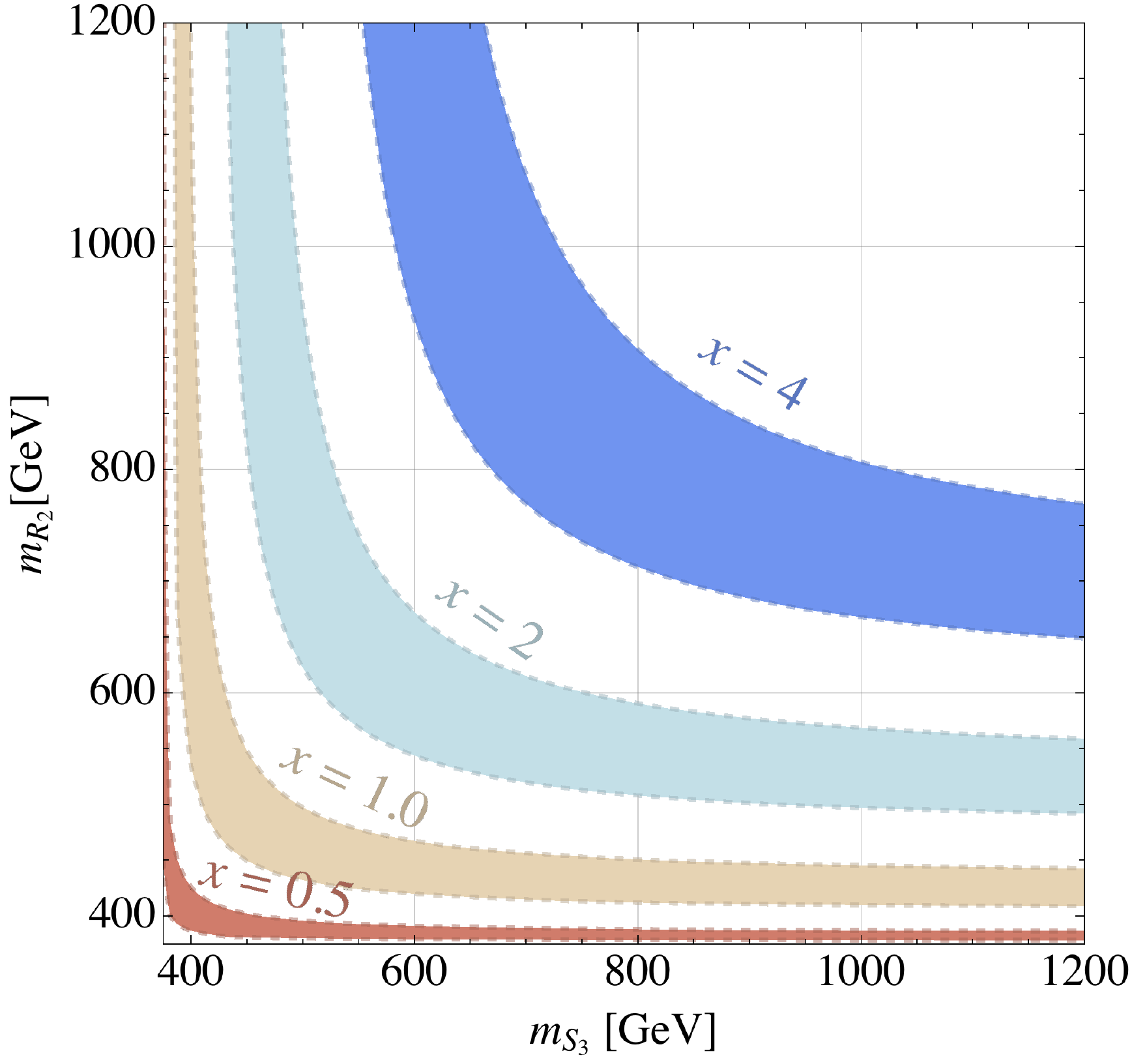}
  \caption{Parameter space of the scenario that satisfies the constraint on $\sigma (pp \to \Sigma_0 \to \gamma\gamma) \in [3.5,7]\e{fb}$ in $m_{S_3}$--$m_{R_3}$ plane for different values of parameter $x$.}
  \label{fig:S3_R2}
\end{figure}

It might be the case that both $S_3$ and $R_2$ contribute towards the
diphoton signal. We accordingly present viable parameter space in the
$m_{S_3}$--$m_{R_2}$ plane for four values of $x(=0.5,1,2,4)$ that
yield $\sigma (pp \to \Sigma_0 \to \gamma\gamma) \in [3.5,7]\e{fb}$ in
Fig.~\ref{fig:S3_R2}. We stress again that it is a prediction
of $SU(5)$ symmetry that parameter $x$ is the same for both $R_2$ and $S_3$ states.

%%%%%% %%%%%%%%% 

The lightness of either $S_3$ or $R_2$ could potentially be in tension with the direct search limits. Both scalars have correct quantum numbers to be leptoquarks (LQs)~\cite{Buchmuller:1986zs}. $S_3$ could, furthermore, mediate proton decay if all possible couplings with the SM fermions are present. We now demonstrate viability of our proposal with a special emphasis on the fermion mass generation within $SU(5)$. 

The couplings of $S_3$ and $R_2$ in the $45$-dimensional representation with the SM fermions that reside in the $10$- and $5$-dimensional representations of $SU(5)$ originate from two contractions in the $SU(5)$ space. These are $Y^{45}_{1\,\alpha \beta} \mathbf{10}_\alpha \mathbf{10}_\beta \mathbf{45}$ and $Y^{45}_{2\,\alpha \beta} \mathbf{10}_\alpha \overline{\mathbf{5}}_\beta \mathbf{45}^*$, where $\mathbf{10}_\alpha \equiv (\mathbf{1},\mathbf{1},1)_\alpha \oplus(\overline{\mathbf{3}},\mathbf{1},-2/3)_\alpha
\oplus(\mathbf{3},\mathbf{2},1/6)_\alpha=(e^C_\alpha,u^C_\alpha,Q_\alpha)$ and
$\overline{\mathbf{5}}_\beta \equiv (\mathbf{1},\mathbf{2},-1/2)_\beta \oplus
(\overline{\mathbf{3}},\mathbf{1},1/3)_\beta=(L_\beta,d^C_\beta)$~\cite{Georgi:1974sy}. The elements of Yukawa coupling matrices are denoted with $Y^{45}_{1\,\alpha \beta}$ and $Y^{45}_{2\,\alpha \beta}$, where $\alpha, \beta=1,2,3$ are flavor indices. The $SU(5)$ indices, on the other hand, are suppressed for clarity. One must also introduce one $5$-dimensional scalar representation ($\mathbf{5}$) if one wants to generate viable masses of the SM charged fermions through VEVs of electrically neutral Higgs-like fields in $\mathbf{5}$ and $\mathbf{45}$~\cite{Georgi:1979df}. The relevant contractions are $Y^{5}_{1\,\alpha \beta} \mathbf{10}_\alpha \mathbf{10}_\beta \mathbf{5}$ and $Y^{5}_{2\,\alpha \beta} \mathbf{10}_\alpha \overline{\mathbf{5}}_\beta \mathbf{5}^*$. We denote VEVs of $\mathbf{5} \equiv \mathbf{5}^i$ and $\mathbf{45} \equiv \mathbf{45}^{ij}_k$ with $\langle\mathbf{5}^5\rangle= v_5 / \sqrt{2}$ and $\langle\mathbf{45}^{15}_{1}\rangle= \langle\mathbf{45}^{2 5}_{2}\rangle=\langle\mathbf{45}^{3 5}_{3}\rangle = v_{45}/\sqrt{2}$, where the $SU(5)$ indices are shown for clarity. The mass matrices of the SM charged fermions are
\begin{align}
\label{eq:m_D}
m_D&=- Y^{45}_2 v_{45}- Y^{5}_2 v_{5}/2,\\
\label{eq:m_E}
m_E&=3 Y^{45\,T}_2 v_{45}- Y^{5\,T}_2 v_{5}/2,\\
\label{eq:m_U}
m_U&=2 \sqrt{2}(Y^{45}_1-Y^{45\,T}_1)v_{45} - \sqrt{2}(Y^{5}_1+Y^{5\,T}_1) v_{5},
\end{align}
where the VEVs are taken to be real. The VEV normalization yields $v_5^2/2+12 v_{45}^2=v^2$, where $v(=246\,\mathrm{GeV})$ is the electroweak VEV~\cite{Dorsner:2011ai}. $m_D$, $m_E$, and $m_U$ are $3 \times 3$ mass matrices for down-type quarks, up-type quarks, and charged leptons in flavor basis, respectively.

%%%%%% %%%%%%%%% 

Let us first assume that the only operators present are the ones that
are needed to generate viable masses of the SM charged
fermions~\cite{Georgi:1979df}. These operators are proportional to
$Y^{5}_1$, $Y^{5}_2$, and $Y^{45}_2$. The couplings of both $R_2$ and
$S_3$ that originate from contraction $Y^{45}_{2\,\alpha \beta}
\mathbf{10}_\alpha \overline{\mathbf{5}}_\beta \mathbf{45}^*$ are of
the leptoquark nature. (See Table II~of Ref.~\cite{Dorsner:2012nq} for
explicit evaluation of the aforementioned contraction with regard to
the $S_3$ couplings.) Note that $S_3$ can only couple with the quark
doublet and the leptonic doublet in this particular instance in the
gauge invariant way. Moreover, the requirement to have experimentally
viable masses for the SM fermions predicts prompt decays of $S_3$ and
$R_2$ if and when these are produced at the LHC. To demonstrate that
prediction it is sufficient to eliminate $Y^{5}_2$ from Eqs.~\eqref{eq:m_D} and \eqref{eq:m_E}. The GUT scale relation $m^T_E-m_D=4Y^{45}_2 v_{45}$ and the fact that $2 m_b(m_{\mathrm{GUT}}) \approx m_\tau(m_{\mathrm{GUT}}) = 1.56$\,GeV~\cite{Dorsner:2011ai} enable one to establish a lower bound on the largest matrix element in $|Y^{45}_2|$. If we take the limit $v_5 \rightarrow 0$ we find that the largest entry of $|Y^{45}_2|$ exceeds $2 \times 10^{-4}$. ($m_b(m_{\mathrm{GUT}})$ and $m_\tau(m_{\mathrm{GUT}})$ are masses of bottom quark and $\tau$ lepton at the GUT scale, respectively. In non-supersymmetric setting the running of these masses does not depend strongly on the threshold corrections.) Note that this is conservative bound since $v_5$ cannot be too small in order to produce top quark mass through Eq.~\eqref{eq:m_U}. This result implies that the direct searches for the scalar LQ states at the LHC are applicable to this particular scenario.

The lower mass limits on $R_2$ and $S_3$ within this particular {\it ansatz} thus originate from three complementary types of experimental searches for leptoquarks at the LHC. The most stringent limits originate from (i) a search for pair production of first generation LQs~\cite{Khachatryan:2015vaa,Aad:2015caa}, (ii) a search for pair production of second generation LQs~\cite{Khachatryan:2015vaa,Aad:2015caa}, and (iii) a search for pair production of third generation scalar LQs~\cite{Khachatryan:2014ura,Khachatryan:2015bsa}. The most relevant bounds from these searches are all based on the data sets collected at the LHC in proton-proton collisions at the center-of-mass energy of $\sqrt{s} = 8$\,TeV. The most constraining of these lower bounds is the one on the mass of second generation LQs that is at $1080$\,GeV, where the branching fraction of LQ to decay into a charged lepton--quark pair is taken to be equal to one. We will concentrate on $R_2$ and argue that $m_{R_2}$ can actually be as low as $400$\,GeV and still be experimentally allowed. For the other leptoquark we will take that the most conservative experimental bound applies to simplify discussion.

The operator $Y^{45}_{2\,\alpha \beta} \mathbf{10}_\alpha \overline{\mathbf{5}}_\beta \mathbf{45}^*$ implies that $R_2$ couples to the right-handed up-type quarks and the leptonic doublets. More specifically, the $R_2$ component with the electric charge of $5/3$ ($2/3$) couples to the right-handed up-type quarks and charged (neutral) leptons. Let us explicitly assume that the $R_2$ component with the $5/3$ charge decays 50\% of the time into the top--$\tau$ lepton pairs and 50\% of the time into the top--$\mu$ pairs. The $R_2$ component with $2/3$ charge then decays 100\% into a top--$\nu$ final state. (One needs to perform summation over $\nu$'s in the final state.) There are no stringent constraints from the LHC on the top--$\nu$ decays for the pair production of leptoquarks. And, the fact that branching fraction is $0.5$ for going into the top--$\tau$ lepton pairs tells us that the LHC bound on the $R_2$ component with the $5/3$ charge is roughly $500$\,GeV. See right panel of Fig.~3 in Ref.~\cite{Dumont:2016xpj} for the implications of the recast of the search for LQs that decay into the top--$\tau$ lepton pairs that was performed by the CMS Collaboration~\cite{Khachatryan:2015bsa}. We can make the $5/3$ component of $R_2$ as light as $400$\,GeV if needed by allowing additional decays into the top--$e$ pairs. Note that it is reasonable to assume that the entries of $Y^{45}_2$ that are associated with the third generation dominate. In our case we assume that the third row of $Y^{45}_2$ is the dominant one to demonstrate that $R_2$ can be very light. This {\it ansatz} is also stable under the renormalization group equation (RGE) running from the GUT scale down to the electroweak scale. The stability under the RGE running when one column (or row) in the Yukawa matrix that determines the LQ couplings to the SM fermions dominates has been demonstrated in Ref.~\cite{Wise:2014oea}.

In view of the preceding discussion we choose to vary $m_{R_2}$ from
$400$ to $700$\,GeV in the right panel of Fig.~\ref{fig:S3R2}. (Scalar LQ multiplets that transform differently with regard to the SM gauge group have also been proposed to help accommodate diphoton excess in Refs.~\cite{Bauer:2015boy,Hati:2016thk}. Viable scenarios with vector LQs have been presented in Refs.~\cite{Murphy:2015kag,deBlas:2015hlv}.)

The second case we discuss is that the only operators present are the ones proportional to $Y^{5}_1$, $Y^{5}_2$, and $Y^{45}_1$. The most important feature of the contraction $Y^{45}_{1\,\alpha \beta} \mathbf{10}_\alpha \mathbf{10}_\beta \mathbf{45}$ is that $S_3$ has only the ``diquark'' couplings with the SM fermions in the $10$-dimensional representation. This is easy to understand since an $SU(2)$ triplet like $S_3$ should couple to a pair of the $SU(2)$ quark doublets in order to create an invariant operator under $SU(3) \times SU(2) \times U(1)$ since the leptonic doublet is not at the disposal. (See Table II~of Ref.~\cite{Dorsner:2012nq} for explicit evaluation of the aforementioned contraction with regard to the $S_3$ couplings.) $SU(5)$ gauge group also dictates that the $S_3$ couplings to quarks are antisymmetric in flavor space. We accordingly require that $Y^{45}_1$ is not a symmetric matrix in order to insure that $S_3$ is coupled to quarks. Note that the issue of the mismatch between the down-type quarks and the charged leptons is not addressed in this particular instance. To do that one would need to introduce, for example, additional vector-like representations.  

The most current constraints that are relevant for the allowed mass of
the $S_3$ multiplet components, if $S_3$ is of ``diquark'' nature,
originate from a search for pair-produced resonances decaying to jet
pairs in $pp$ collisions at the LHC~\cite{Khachatryan:2014lpa}. We
conservatively interpret these measurements to imply lower limit on
the mass of the $S_3$ ``diquark'' to be at $390$\,GeV. This is thus
adopted as the lowest value of parameter $m_{S_3}$ that we use to
present our results in the left panel of Fig.~\ref{fig:S3R2}. This time around $R_2$ couples to the right-handed charged leptons and the quark doublets. More specifically, the $R_2$ component with the $5/3$ ($2/3$) charge couples to the right-handed charged leptons and up-type (down-type) quarks. In this instance the most conservative experimental limit for the mass of $R_2$ leptoquark holds true.

%%%%%% %%%%%%%%% 

The last scenario, and the most general one, that we want to address
is when all four operators that contribute towards charged fermion
masses are present. In this case $S_3$ has both ``diquark'' and
leptoquark couplings~\cite{Dorsner:2012nq}. This simply means that the
proton decay constraints stipulate that $S_3$ cannot contribute
towards the diphoton signal. This particular scenario corresponds to
the $R_2$ dominance that is shown in the right panel of Fig.~\ref{fig:S3R2} and the predictions in the third row of Table~\ref{tab:dib}. Note that it is sufficient that the entries of $Y^{45}_2$ dominate over entries in $Y^{45}_1$, where $Y^{45}_2$ has a form that predominantly couples the $R_2$ component with the $5/3$ charge to the right-handed top quark and charged leptons. 

%%%%%%%UNIFICATION-BEGINNING%%%%%%%

Let us finally address the issue of unification of gauge couplings within the non-supersymmetric $SU(5)$ framework with $5$- and $45$-dimensional scalar representations. To do that we first define quantities $b^J_{ij}=(b^J_{i}-b^J_{j})$, $i,j=1,2,3$, where $b^J_{i}$ are the $\beta$-function coefficients of particle $J$ with mass $m_J$. $b^J_{1}$, $b^J_{2}$, and $b^J_{3}$ are associated with $U(1)$, $SU(2)$, and $SU(3)$ of the SM, respectively. We furthermore introduce coefficients $B_{ij}=\sum_{J} b_{ij}^J r_{J}$, where the sum goes through all particles that reside below the GUT scale and parameter $r_J=(\ln
m_{\mathrm{GUT}}/m_{J})/(\ln m_{\mathrm{GUT}}/m_{Z})$ describes where between $Z$ boson mass and the GUT scale particle $J$ is.   

The gauge coupling at the GUT scale $\alpha_{\mathrm{GUT}}$ is well-behaved in non-supersymmetric $SU(5)$ framework and it can, accordingly, be eliminated using three equations that describe running of individual gauge couplings below the GUT scale. This leaves two relevant equations that read~\cite{Giveon:1991zm}
\begin{eqnarray}
\label{eq:a}
\frac{B_{23}}{B_{12}}&=&\frac{5}{8}
\frac{\sin^2
\theta_W-\alpha/\alpha_S}{3/8-\sin^2 \theta_W}=0.721 \pm 0.004,\\
\label{eq:b}
\ln \frac{m_{\mathrm{GUT}}}{m_Z}&=&\frac{16 \pi}{5
\alpha} \frac{3/8-\sin^2 \theta_W}{B_{12}}=\frac{184.8 \pm
0.1}{B_{12}},
\end{eqnarray}
where we use $\alpha_S(m_Z)=0.1193\pm0.0016$, $\alpha^{-1}(m_Z)=127.906\pm0.019$, and
$\sin^2 \theta_W=0.23126\pm0.00005$~\cite{Agashe:2014kda} to produce numerical values in the right-hand sides of Eqs.~\eqref{eq:a} and \eqref{eq:b}. Eq.~\eqref{eq:a}, if satisfied, insures that the gauge couplings meet whereas Eq.~\eqref{eq:b} provides the corresponding value of the GUT scale. 

The SM content yields $B^{\textrm{SM}}_{23}/B^{\textrm{SM}}_{12}=0.53$ instead of the experimentally required value given in Eq.~\eqref{eq:a}. Ideally, one would like to have a light field $J$ with positive $b^J_{23}$ and negative $b^J_{12}$. This would not only help in bringing the left-hand side of Eq.~\eqref{eq:a} in agreement with the required experimental value but would also raise the GUT scale $m_{\mathrm{GUT}}$ through Eq.~\eqref{eq:b}. As it turns out, $S_3$ is an ideal candidate with $b^{S_3}_{23}=9/6$ and $b^{S_3}_{12}=-27/15$. The corresponding coefficients of leptoquark $R_2$ are $b^{R_2}_{23}=1/6$ and $b^{R_2}_{12}=17/15$. We find that unification is possible for light $S_3$ and $R_2$. For example, if we set $m_{S_3}=400$\,GeV and $m_{R_2}=2$\,TeV we obtain exact unification for central values of input parameters with an upper bound on the GUT scale that is $m_{\mathrm{GUT}}\leq 6 \times 10^{15}$\,GeV. The particle content comprises three scalar representations, i.e., $5$-, $24$-, and $45$-dimensional representations of $SU(5)$, one adjoint representation with the gauge fields, and the SM fermions. Unification is obtained assuming that all proton decay mediating scalars are at or above $10^{12}$\,GeV. (The maximal value of the GUT scale grows with the mass of $R_2$ leptoquark.) This demonstrates that light $S_3$ and/or $R_2$ represent viable options within non-supersymmetric $SU(5)$ framework with $5$- and $45$-dimensional scalar representations.

%%%%%%%UNIFICATION-END%%%%%%%

Our proposal opens up a possibility to have one light SM singlet at the electroweak scale in practically any $SU(5)$ setting without the need to introduce {\it ad hoc} scalars. We furthermore demonstrate viability of our proposal using individual or combined contributions towards diphoton signal of scalar multiplets that transform as $(\mathbf{3},\mathbf{3},-1/3)$ and $(\mathbf{3},\mathbf{2},7/6)$ under the SM gauge group. We relate the existence of these colored scalars to the issue of fermion mass generation in $SU(5)$ and provide predictions for the diboson decays of the scalar singlet state at the LHC.     

%%%%%%%%%%%%%%%%%%%%%%%%%%%%%%%%%%%%%%%%%%%%%%%%%%

\acknowledgments
This work has been supported in part by Croatian Science Foundation
under the project 7118. I.D.\ thanks A.\ Greljo for insightful
discussions. S.F.\ and N.K.\ thank J.\ F.\ Kamenik for informative
discussions. This work has been supported in part by the Slovenian Research Agency.

%%%%%%%%%%%%%%%%%%%%%%%%%%%%%%%%%%%%%%%%%%%%%%%%%%

\appendix

\section{Diboson decay amplitudes}
\label{sec:appendix} 
The decay amplitude of a scalar resonance to diboson final
states, $\Sigma_0(q) \to
V(p,\epsilon) V' (p',\epsilon')$, can be expressed in terms of two
form factors
\begin{equation}
  \label{eq:AmpVV}
  \mc{A}_{\Sigma_0 \to V V^\prime} = \frac{-i m_{\Sigma_0}}{2\pi}
  \left[A_{VV^\prime} g^{\mu\nu} - 2 B_{VV^\prime} \frac{p^{\prime \mu} p^\nu}{m_{\Sigma_0}^2} \right]
  \epsilon^\ast_\mu \epsilon^{\prime\ast}_\nu .
\end{equation}
Ward identity states that the amplitude ~\eqref{eq:AmpVV} vanishes
whenever we replace external polarization of a photon or a gluon with
its momentum, and this requires that form factors $A_{VV^\prime}$ and
$B_{VV^\prime}$ are not independent. Notice that transversality
conditions, $\epsilon\cdot p = \epsilon^\prime \cdot p^\prime=0$,
allow replacing $p^{\prime \mu} p^\nu$ by $q^\mu q^\nu$ in
Eq.~\eqref{eq:AmpVV}, however, in this case one has to enforce
transversality also in the polarization sum prescription:
$\sum_\lambda \epsilon_\mu(p,\lambda) \epsilon^*_\nu(p,\lambda) \to
-g_{\mu\nu} + \frac{p_\mu p_\nu}{p^2}$,
regardless of whether vector $\epsilon(p,\lambda)$ is massless or
not. For each diboson decay amplitude mediated by the $S_3$
state the form factors $A_{VV^\prime}$, $B_{VV^\prime}$, that we
present below, have been reduced to the Passarino-Veltman functions
with the help of FeynCalc~\cite{Shtabovenko:2016sxi,Mertig:1990an} and
numerically evaluated using the LoopTools package~\cite{Hahn:1998yk}. In
the following expressions, gluon indices are denoted by
$A,B(=1,\ldots,8)$, and one has to insert the quantum numbers of
$S_3$, i.e., $T=1$, $Y=-1/3$. Weak mixing
factors $\tan \theta_W$ and $\sin \theta_W$
($\sin^2 \theta_W = 0.231$) are abbreviated as $t_\theta$ and
$s_\theta$, respectively. The amplitudes for the cases that involve a
massless boson in the final state read:
\begin{align}
%%%%%
%%%%% \gamma\gamma
%%%%%
A_{g^A g^B} &= B_{g^A g^B} = x \alpha_S \frac{\delta^{AB}}{2} (2T+1) \Big[ 1+ 2 m_{\Sigma_0}^2
  C_0(0,m_{\Sigma_0}^2,0,m_{S_3}^2,m_{S_3}^2,m_{S_3}^2)\Big],\\
A_{\gamma\gamma} &= B_{\gamma\gamma} = x \alpha N_c (2T+1)\left[
  Y^2+\frac{T(T+1)}{3}\right]  \Big[ 1+ 2 m_{\Sigma_0}^2
  C_0(0,m_{\Sigma_0}^2,0,m_{S_3}^2,m_{S_3}^2,m_{S_3}^2)\Big],\\
%%%%%
%%%%% Z\gamma
%%%%%
A_{Z\gamma} &= \frac{B_{Z\gamma}}{1-\frac{m_Z^2}{m_{\Sigma_0}^2}}\\
 & = x \frac{\alpha N_c}{t_{\theta}} (2T+1)\left[-Y^2
  t^2_\theta + \frac{T(T+1)}{3} \right]
\Bigg\{1+\frac{m_Z^2\left[B_0(m_Z^2,m_{S_3}^2,m_{S_3}^2) - B_0(m_{\Sigma _0}^2,m_{S_3}^2,m_{S_3}^2)\right]}{m_Z^2-m_{\Sigma_0}^2}\nn\\
&\quad+m_{S_3}^2 \left[C_0(0,m_{\Sigma
   _0}^2,m_Z^2,m_{S_3}^2,m_{S_3}^2,m_{S_3}^2)+C_0(m_Z^2,m_{\Sigma
   _0}^2,0,m_{S_3}^2,m_{S_3}^2,m_{S_3}^2)\right]\Bigg\}\nn ,
\end{align}
whereas for the massive final states one finds:
\begin{align}
%%%%%
%%%%% ZZ
%%%%%
A_{ZZ} &= x \frac{\alpha N_c}{t_\theta^2 }  (2T+1)\left[Y^2
  t_\theta^4+\frac{T(T+1)}{3} \right]        \\
&\quad\times\Bigg\{1 + \frac{2 m_Z^2 \left[B_0(m_{\Sigma_0}^2,m_{S_3}^2,m_{S_3}^2)-B_0(m_Z^2,m_{S_3}^2,m_{S_3}^2)\right]}{m_{\Sigma _0}^2-4 m_Z^2}\nn
\\&\quad
+2\left[m_{S_3}^2 + \frac{m_Z^4}{m_{\Sigma _0}^2-4m_Z^2}\right] C_0(m_Z^2,m_{\Sigma
   _0}^2,m_Z^2,m_{S_3}^2,m_{S_3}^2,m_{S_3}^2)\Bigg\}\nn ,\\
B_{ZZ} &= x \frac{\alpha N_c}{t_{\theta }^2}   (2T+1)\left[Y^2
  t_\theta^4+\frac{T(T+1)}{3} \right]\frac{1}{m_{\Sigma _0}^2-4 m_Z^2}\\
&\quad\times\Bigg\{m_{\Sigma _0}^2-2 m_Z^2-\frac{2 m_Z^2 \left(m_{\Sigma _0}^2+2 m_Z^2\right)
   \left[B_0(m_Z^2,m_{S_3}^2,m_{S_3}^2)-B_0(m_{\Sigma
         _0}^2,m_{S_3}^2,m_{S_3}^2)\right]}{m_{\Sigma _0}^2-4 m_Z^2}\nn\\
&\quad+2 \left[m_{S_3}^2 \left(m_{\Sigma _0}^2-2
   m_Z^2\right)+2 m_Z^4 \left(1+\frac{3m_Z^2}{m_{\Sigma_0}^2-4m_Z^2} \right)\right] C_0(m_Z^2,m_{\Sigma
   _0}^2,m_Z^2,m_{S_3}^2,m_{S_3}^2,m_{S_3}^2)\Bigg\}\nn ,\\
%%%%%
%%%%% WW
%%%%%
A_{WW} &= x \frac{2 \alpha  N_c}{s_{\theta }^2} \Bigg\{1+
\frac{2m_W^2\left[B_0(m_{\Sigma_0}^2,m_{S_3}^2,m_{S_3}^2)-B_0(m_W^2,m_{S_3}^2,m_{S_3}^2)\right]}{m_{\Sigma
    _0}^2-4 m_W^2}\\
&\quad+ 2\left(m_{S_3}^2+\frac{m_W^4}{m_{\Sigma _0}^2-4 m_W^2}\right)
C_0(m_W^2,m_{\Sigma
    _0}^2,m_W^2,m_{S_3}^2,m_{S_3}^2,m_{S_3}^2)\Bigg\}\nn ,\\
B_{WW} &= x \frac{2 \alpha  N_c}{s_{\theta }^2}\frac{1}{\left(m_{\Sigma
      _0}^2-4 m_W^2\right)}\\
&\quad\times\Bigg\{m_{\Sigma _0}^2-2 m_W^2 -\frac{2 m_W^2 \left(m_{\Sigma _0}^2+2 m_W^2\right) \left[B_0(m_W^2,m_{S_3}^2,m_{S_3}^2)-B_0(m_{\Sigma
   _0}^2,m_{S_3}^2,m_{S_3}^2)\right]}{m_{\Sigma
      _0}^2-4 m_W^2}\nn\\
&\quad+2 \left[m_{S_3}^2 \left(m_{\Sigma _0}^2-2 m_W^2\right)+\frac{2 m_W^4 (m_{\Sigma _0}^2 -
   m_W^2)}{m_{\Sigma _0}^2-4 m_W^2}\right] C_0(m_W^2,m_{\Sigma _0}^2,m_W^2,m_{S_3}^2,m_{S_3}^2,m_{S_3}^2)\Bigg\}\nn.
  \end{align}
The amplitudes due to virtual $R_2$ contributions are obtained by
adjusting the mass $m_{S_3} \to m_{R_2}$, inserting
appropriate values for $T$ and $Y$ for the electrically neutral final
state amplitudes, and adjusting the $WW$ amplitudes by a factor
of $1/4$.

\bibliography{references}

\begin{thebibliography}{39}
\expandafter\ifx\csname natexlab\endcsname\relax\def\natexlab#1{#1}\fi
\expandafter\ifx\csname bibnamefont\endcsname\relax
  \def\bibnamefont#1{#1}\fi
\expandafter\ifx\csname bibfnamefont\endcsname\relax
  \def\bibfnamefont#1{#1}\fi
\expandafter\ifx\csname citenamefont\endcsname\relax
  \def\citenamefont#1{#1}\fi
\expandafter\ifx\csname url\endcsname\relax
  \def\url#1{\texttt{#1}}\fi
\expandafter\ifx\csname urlprefix\endcsname\relax\def\urlprefix{URL }\fi
\providecommand{\bibinfo}[2]{#2}
\providecommand{\eprint}[2][]{\url{#2}}

\bibitem[{\citenamefont{Aaboud et~al.}(2016)}]{Aaboud:2016tru}
\bibinfo{author}{\bibfnamefont{M.}~\bibnamefont{Aaboud}} \bibnamefont{et~al.}
  (\bibinfo{collaboration}{ATLAS}) (\bibinfo{year}{2016}), \eprint{1606.03833}.

\bibitem[{\citenamefont{Khachatryan
  et~al.}(2016{\natexlab{a}})}]{Khachatryan:2016hje}
\bibinfo{author}{\bibfnamefont{V.}~\bibnamefont{Khachatryan}}
  \bibnamefont{et~al.} (\bibinfo{collaboration}{CMS})
  (\bibinfo{year}{2016}{\natexlab{a}}), \eprint{1606.04093}.

\bibitem[{\citenamefont{Knapen et~al.}(2016)\citenamefont{Knapen, Melia,
  Papucci, and Zurek}}]{Knapen:2015dap}
\bibinfo{author}{\bibfnamefont{S.}~\bibnamefont{Knapen}},
  \bibinfo{author}{\bibfnamefont{T.}~\bibnamefont{Melia}},
  \bibinfo{author}{\bibfnamefont{M.}~\bibnamefont{Papucci}}, \bibnamefont{and}
  \bibinfo{author}{\bibfnamefont{K.}~\bibnamefont{Zurek}},
  \bibinfo{journal}{Phys. Rev.} \textbf{\bibinfo{volume}{D93}},
  \bibinfo{pages}{075020} (\bibinfo{year}{2016}), \eprint{1512.04928}.

\bibitem[{\citenamefont{Buttazzo et~al.}(2016)\citenamefont{Buttazzo, Greljo,
  and Marzocca}}]{Buttazzo:2015txu}
\bibinfo{author}{\bibfnamefont{D.}~\bibnamefont{Buttazzo}},
  \bibinfo{author}{\bibfnamefont{A.}~\bibnamefont{Greljo}}, \bibnamefont{and}
  \bibinfo{author}{\bibfnamefont{D.}~\bibnamefont{Marzocca}},
  \bibinfo{journal}{Eur. Phys. J.} \textbf{\bibinfo{volume}{C76}},
  \bibinfo{pages}{116} (\bibinfo{year}{2016}), \eprint{1512.04929}.

\bibitem[{\citenamefont{Franceschini
  et~al.}(2016{\natexlab{a}})\citenamefont{Franceschini, Giudice, Kamenik,
  McCullough, Pomarol, Rattazzi, Redi, Riva, Strumia, and
  Torre}}]{Franceschini:2015kwy}
\bibinfo{author}{\bibfnamefont{R.}~\bibnamefont{Franceschini}},
  \bibinfo{author}{\bibfnamefont{G.~F.} \bibnamefont{Giudice}},
  \bibinfo{author}{\bibfnamefont{J.~F.} \bibnamefont{Kamenik}},
  \bibinfo{author}{\bibfnamefont{M.}~\bibnamefont{McCullough}},
  \bibinfo{author}{\bibfnamefont{A.}~\bibnamefont{Pomarol}},
  \bibinfo{author}{\bibfnamefont{R.}~\bibnamefont{Rattazzi}},
  \bibinfo{author}{\bibfnamefont{M.}~\bibnamefont{Redi}},
  \bibinfo{author}{\bibfnamefont{F.}~\bibnamefont{Riva}},
  \bibinfo{author}{\bibfnamefont{A.}~\bibnamefont{Strumia}}, \bibnamefont{and}
  \bibinfo{author}{\bibfnamefont{R.}~\bibnamefont{Torre}},
  \bibinfo{journal}{JHEP} \textbf{\bibinfo{volume}{03}}, \bibinfo{pages}{144}
  (\bibinfo{year}{2016}{\natexlab{a}}), \eprint{1512.04933}.

\bibitem[{\citenamefont{Georgi and Glashow}(1974)}]{Georgi:1974sy}
\bibinfo{author}{\bibfnamefont{H.}~\bibnamefont{Georgi}} \bibnamefont{and}
  \bibinfo{author}{\bibfnamefont{S.~L.} \bibnamefont{Glashow}},
  \bibinfo{journal}{Phys. Rev. Lett.} \textbf{\bibinfo{volume}{32}},
  \bibinfo{pages}{438} (\bibinfo{year}{1974}).

\bibitem[{\citenamefont{Guth and Weinberg}(1981)}]{Guth:1981uk}
\bibinfo{author}{\bibfnamefont{A.~H.} \bibnamefont{Guth}} \bibnamefont{and}
  \bibinfo{author}{\bibfnamefont{E.~J.} \bibnamefont{Weinberg}},
  \bibinfo{journal}{Phys. Rev.} \textbf{\bibinfo{volume}{D23}},
  \bibinfo{pages}{876} (\bibinfo{year}{1981}).

\bibitem[{\citenamefont{Witten}(1980)}]{Witten:1979nr}
\bibinfo{author}{\bibfnamefont{E.}~\bibnamefont{Witten}},
  \bibinfo{journal}{Phys. Lett.} \textbf{\bibinfo{volume}{B91}},
  \bibinfo{pages}{81} (\bibinfo{year}{1980}).

\bibitem[{\citenamefont{Patel and Sharma}(2016)}]{Patel:2015ulo}
\bibinfo{author}{\bibfnamefont{K.~M.} \bibnamefont{Patel}} \bibnamefont{and}
  \bibinfo{author}{\bibfnamefont{P.}~\bibnamefont{Sharma}},
  \bibinfo{journal}{Phys. Lett.} \textbf{\bibinfo{volume}{B757}},
  \bibinfo{pages}{282} (\bibinfo{year}{2016}), \eprint{1512.07468}.

\bibitem[{\citenamefont{Hall et~al.}(2016)\citenamefont{Hall, Harigaya, and
  Nomura}}]{Hall:2015xds}
\bibinfo{author}{\bibfnamefont{L.~J.} \bibnamefont{Hall}},
  \bibinfo{author}{\bibfnamefont{K.}~\bibnamefont{Harigaya}}, \bibnamefont{and}
  \bibinfo{author}{\bibfnamefont{Y.}~\bibnamefont{Nomura}},
  \bibinfo{journal}{JHEP} \textbf{\bibinfo{volume}{03}}, \bibinfo{pages}{017}
  (\bibinfo{year}{2016}), \eprint{1512.07904}.

\bibitem[{\citenamefont{Dutta et~al.}(2016)\citenamefont{Dutta, Gao, Ghosh,
  Gogoladze, Li, Shafi, and Walker}}]{Dutta:2016jqn}
\bibinfo{author}{\bibfnamefont{B.}~\bibnamefont{Dutta}},
  \bibinfo{author}{\bibfnamefont{Y.}~\bibnamefont{Gao}},
  \bibinfo{author}{\bibfnamefont{T.}~\bibnamefont{Ghosh}},
  \bibinfo{author}{\bibfnamefont{I.}~\bibnamefont{Gogoladze}},
  \bibinfo{author}{\bibfnamefont{T.}~\bibnamefont{Li}},
  \bibinfo{author}{\bibfnamefont{Q.}~\bibnamefont{Shafi}}, \bibnamefont{and}
  \bibinfo{author}{\bibfnamefont{J.~W.} \bibnamefont{Walker}}
  (\bibinfo{year}{2016}), \eprint{1601.00866}.

\bibitem[{\citenamefont{Ellis et~al.}(2016)\citenamefont{Ellis, Ellis,
  Quevillon, Sanz, and You}}]{Ellis:2015oso}
\bibinfo{author}{\bibfnamefont{J.}~\bibnamefont{Ellis}},
  \bibinfo{author}{\bibfnamefont{S.~A.~R.} \bibnamefont{Ellis}},
  \bibinfo{author}{\bibfnamefont{J.}~\bibnamefont{Quevillon}},
  \bibinfo{author}{\bibfnamefont{V.}~\bibnamefont{Sanz}}, \bibnamefont{and}
  \bibinfo{author}{\bibfnamefont{T.}~\bibnamefont{You}},
  \bibinfo{journal}{JHEP} \textbf{\bibinfo{volume}{03}}, \bibinfo{pages}{176}
  (\bibinfo{year}{2016}), \eprint{1512.05327}.

\bibitem[{\citenamefont{Buchmuller et~al.}(1987)\citenamefont{Buchmuller,
  Ruckl, and Wyler}}]{Buchmuller:1986zs}
\bibinfo{author}{\bibfnamefont{W.}~\bibnamefont{Buchmuller}},
  \bibinfo{author}{\bibfnamefont{R.}~\bibnamefont{Ruckl}}, \bibnamefont{and}
  \bibinfo{author}{\bibfnamefont{D.}~\bibnamefont{Wyler}},
  \bibinfo{journal}{Phys. Lett.} \textbf{\bibinfo{volume}{B191}},
  \bibinfo{pages}{442} (\bibinfo{year}{1987}), \bibinfo{note}{[Erratum: Phys.
  Lett.B448,320(1999)]}.

\bibitem[{\citenamefont{Djouadi}(2008)}]{Djouadi:2005gj}
\bibinfo{author}{\bibfnamefont{A.}~\bibnamefont{Djouadi}},
  \bibinfo{journal}{Phys. Rept.} \textbf{\bibinfo{volume}{459}},
  \bibinfo{pages}{1} (\bibinfo{year}{2008}), \eprint{hep-ph/0503173}.

\bibitem[{\citenamefont{Dorsner
  et~al.}(2012{\natexlab{a}})\citenamefont{Dorsner, Fajfer, Greljo, and
  Kamenik}}]{Dorsner:2012pp}
\bibinfo{author}{\bibfnamefont{I.}~\bibnamefont{Dorsner}},
  \bibinfo{author}{\bibfnamefont{S.}~\bibnamefont{Fajfer}},
  \bibinfo{author}{\bibfnamefont{A.}~\bibnamefont{Greljo}}, \bibnamefont{and}
  \bibinfo{author}{\bibfnamefont{J.~F.} \bibnamefont{Kamenik}},
  \bibinfo{journal}{JHEP} \textbf{\bibinfo{volume}{11}}, \bibinfo{pages}{130}
  (\bibinfo{year}{2012}{\natexlab{a}}), \eprint{1208.1266}.

\bibitem[{\citenamefont{Bauer and Neubert}(2015)}]{Bauer:2015boy}
\bibinfo{author}{\bibfnamefont{M.}~\bibnamefont{Bauer}} \bibnamefont{and}
  \bibinfo{author}{\bibfnamefont{M.}~\bibnamefont{Neubert}}
  (\bibinfo{year}{2015}), \eprint{1512.06828}.

\bibitem[{\citenamefont{Chatrchyan et~al.}(2013)}]{Chatrchyan:2013txa}
\bibinfo{author}{\bibfnamefont{S.}~\bibnamefont{Chatrchyan}}
  \bibnamefont{et~al.} (\bibinfo{collaboration}{CMS}), \bibinfo{journal}{Eur.
  Phys. J.} \textbf{\bibinfo{volume}{C73}}, \bibinfo{pages}{2604}
  (\bibinfo{year}{2013}), \eprint{1304.7498}.

\bibitem[{\citenamefont{Olive et~al.}(2014)}]{Agashe:2014kda}
\bibinfo{author}{\bibfnamefont{K.~A.} \bibnamefont{Olive}} \bibnamefont{et~al.}
  (\bibinfo{collaboration}{Particle Data Group}), \bibinfo{journal}{Chin.
  Phys.} \textbf{\bibinfo{volume}{C38}}, \bibinfo{pages}{090001}
  (\bibinfo{year}{2014}).

\bibitem[{\citenamefont{Franceschini
  et~al.}(2016{\natexlab{b}})\citenamefont{Franceschini, Giudice, Kamenik,
  McCullough, Riva, Strumia, and Torre}}]{Franceschini:2016gxv}
\bibinfo{author}{\bibfnamefont{R.}~\bibnamefont{Franceschini}},
  \bibinfo{author}{\bibfnamefont{G.~F.} \bibnamefont{Giudice}},
  \bibinfo{author}{\bibfnamefont{J.~F.} \bibnamefont{Kamenik}},
  \bibinfo{author}{\bibfnamefont{M.}~\bibnamefont{McCullough}},
  \bibinfo{author}{\bibfnamefont{F.}~\bibnamefont{Riva}},
  \bibinfo{author}{\bibfnamefont{A.}~\bibnamefont{Strumia}}, \bibnamefont{and}
  \bibinfo{author}{\bibfnamefont{R.}~\bibnamefont{Torre}}
  (\bibinfo{year}{2016}{\natexlab{b}}), \eprint{1604.06446}.

\bibitem[{ATL(2016)}]{ATLAS-CONF-2016-018}
\bibinfo{journal}{ATLAS Collaboration, Report No. ATLAS-CONF-2016-018}
  (\bibinfo{year}{2016}), \urlprefix\url{http://cds.cern.ch/record/2141568}.

\bibitem[{CMS(2016)}]{CMS-PAS-EXO-16-018}
\bibinfo{journal}{CMS Collaboration, Report No. CMS-PAS-EXO-16-018}
  (\bibinfo{year}{2016}), \urlprefix\url{https://cds.cern.ch/record/2139899}.

\bibitem[{\citenamefont{Gupta et~al.}(2015)\citenamefont{Gupta, J{\" a}ger,
  Kats, Perez, and Stamou}}]{Gupta:2015zzs}
\bibinfo{author}{\bibfnamefont{R.~S.} \bibnamefont{Gupta}},
  \bibinfo{author}{\bibfnamefont{S.}~\bibnamefont{J{\" a}ger}},
  \bibinfo{author}{\bibfnamefont{Y.}~\bibnamefont{Kats}},
  \bibinfo{author}{\bibfnamefont{G.}~\bibnamefont{Perez}}, \bibnamefont{and}
  \bibinfo{author}{\bibfnamefont{E.}~\bibnamefont{Stamou}}
  (\bibinfo{year}{2015}), \eprint{1512.05332}.

\bibitem[{\citenamefont{Georgi and Jarlskog}(1979)}]{Georgi:1979df}
\bibinfo{author}{\bibfnamefont{H.}~\bibnamefont{Georgi}} \bibnamefont{and}
  \bibinfo{author}{\bibfnamefont{C.}~\bibnamefont{Jarlskog}},
  \bibinfo{journal}{Phys. Lett.} \textbf{\bibinfo{volume}{B86}},
  \bibinfo{pages}{297} (\bibinfo{year}{1979}).

\bibitem[{\citenamefont{Dorsner et~al.}(2011)\citenamefont{Dorsner, Drobnak,
  Fajfer, Kamenik, and Kosnik}}]{Dorsner:2011ai}
\bibinfo{author}{\bibfnamefont{I.}~\bibnamefont{Dorsner}},
  \bibinfo{author}{\bibfnamefont{J.}~\bibnamefont{Drobnak}},
  \bibinfo{author}{\bibfnamefont{S.}~\bibnamefont{Fajfer}},
  \bibinfo{author}{\bibfnamefont{J.~F.} \bibnamefont{Kamenik}},
  \bibnamefont{and} \bibinfo{author}{\bibfnamefont{N.}~\bibnamefont{Kosnik}},
  \bibinfo{journal}{JHEP} \textbf{\bibinfo{volume}{11}}, \bibinfo{pages}{002}
  (\bibinfo{year}{2011}), \eprint{1107.5393}.

\bibitem[{\citenamefont{Dorsner
  et~al.}(2012{\natexlab{b}})\citenamefont{Dorsner, Fajfer, and
  Kosnik}}]{Dorsner:2012nq}
\bibinfo{author}{\bibfnamefont{I.}~\bibnamefont{Dorsner}},
  \bibinfo{author}{\bibfnamefont{S.}~\bibnamefont{Fajfer}}, \bibnamefont{and}
  \bibinfo{author}{\bibfnamefont{N.}~\bibnamefont{Kosnik}},
  \bibinfo{journal}{Phys. Rev.} \textbf{\bibinfo{volume}{D86}},
  \bibinfo{pages}{015013} (\bibinfo{year}{2012}{\natexlab{b}}),
  \eprint{1204.0674}.

\bibitem[{\citenamefont{Khachatryan
  et~al.}(2016{\natexlab{b}})}]{Khachatryan:2015vaa}
\bibinfo{author}{\bibfnamefont{V.}~\bibnamefont{Khachatryan}}
  \bibnamefont{et~al.} (\bibinfo{collaboration}{CMS}), \bibinfo{journal}{Phys.
  Rev.} \textbf{\bibinfo{volume}{D93}}, \bibinfo{pages}{032004}
  (\bibinfo{year}{2016}{\natexlab{b}}), \eprint{1509.03744}.

\bibitem[{\citenamefont{Aad et~al.}(2016)}]{Aad:2015caa}
\bibinfo{author}{\bibfnamefont{G.}~\bibnamefont{Aad}} \bibnamefont{et~al.}
  (\bibinfo{collaboration}{ATLAS}), \bibinfo{journal}{Eur. Phys. J.}
  \textbf{\bibinfo{volume}{C76}}, \bibinfo{pages}{5} (\bibinfo{year}{2016}),
  \eprint{1508.04735}.

\bibitem[{\citenamefont{Khachatryan et~al.}(2014)}]{Khachatryan:2014ura}
\bibinfo{author}{\bibfnamefont{V.}~\bibnamefont{Khachatryan}}
  \bibnamefont{et~al.} (\bibinfo{collaboration}{CMS}), \bibinfo{journal}{Phys.
  Lett.} \textbf{\bibinfo{volume}{B739}}, \bibinfo{pages}{229}
  (\bibinfo{year}{2014}), \eprint{1408.0806}.

\bibitem[{\citenamefont{Khachatryan
  et~al.}(2015{\natexlab{a}})}]{Khachatryan:2015bsa}
\bibinfo{author}{\bibfnamefont{V.}~\bibnamefont{Khachatryan}}
  \bibnamefont{et~al.} (\bibinfo{collaboration}{CMS}), \bibinfo{journal}{JHEP}
  \textbf{\bibinfo{volume}{07}}, \bibinfo{pages}{042}
  (\bibinfo{year}{2015}{\natexlab{a}}), \eprint{1503.09049}.

\bibitem[{\citenamefont{Dumont et~al.}(2016)\citenamefont{Dumont, Nishiwaki,
  and Watanabe}}]{Dumont:2016xpj}
\bibinfo{author}{\bibfnamefont{B.}~\bibnamefont{Dumont}},
  \bibinfo{author}{\bibfnamefont{K.}~\bibnamefont{Nishiwaki}},
  \bibnamefont{and} \bibinfo{author}{\bibfnamefont{R.}~\bibnamefont{Watanabe}}
  (\bibinfo{year}{2016}), \eprint{1603.05248}.

\bibitem[{\citenamefont{Wise and Zhang}(2014)}]{Wise:2014oea}
\bibinfo{author}{\bibfnamefont{M.~B.} \bibnamefont{Wise}} \bibnamefont{and}
  \bibinfo{author}{\bibfnamefont{Y.}~\bibnamefont{Zhang}},
  \bibinfo{journal}{Phys. Rev.} \textbf{\bibinfo{volume}{D90}},
  \bibinfo{pages}{053005} (\bibinfo{year}{2014}), \eprint{1404.4663}.

\bibitem[{\citenamefont{Hati}(2016)}]{Hati:2016thk}
\bibinfo{author}{\bibfnamefont{C.}~\bibnamefont{Hati}}, \bibinfo{journal}{Phys.
  Rev.} \textbf{\bibinfo{volume}{D93}}, \bibinfo{pages}{075002}
  (\bibinfo{year}{2016}), \eprint{1601.02457}.

\bibitem[{\citenamefont{Murphy}(2016)}]{Murphy:2015kag}
\bibinfo{author}{\bibfnamefont{C.~W.} \bibnamefont{Murphy}},
  \bibinfo{journal}{Phys. Lett.} \textbf{\bibinfo{volume}{B757}},
  \bibinfo{pages}{192} (\bibinfo{year}{2016}), \eprint{1512.06976}.

\bibitem[{\citenamefont{de~Blas et~al.}(2015)\citenamefont{de~Blas, Santiago,
  and Vega-Morales}}]{deBlas:2015hlv}
\bibinfo{author}{\bibfnamefont{J.}~\bibnamefont{de~Blas}},
  \bibinfo{author}{\bibfnamefont{J.}~\bibnamefont{Santiago}}, \bibnamefont{and}
  \bibinfo{author}{\bibfnamefont{R.}~\bibnamefont{Vega-Morales}}
  (\bibinfo{year}{2015}), \eprint{1512.07229}.

\bibitem[{\citenamefont{Khachatryan
  et~al.}(2015{\natexlab{b}})}]{Khachatryan:2014lpa}
\bibinfo{author}{\bibfnamefont{V.}~\bibnamefont{Khachatryan}}
  \bibnamefont{et~al.} (\bibinfo{collaboration}{CMS}), \bibinfo{journal}{Phys.
  Lett.} \textbf{\bibinfo{volume}{B747}}, \bibinfo{pages}{98}
  (\bibinfo{year}{2015}{\natexlab{b}}), \eprint{1412.7706}.

\bibitem[{\citenamefont{Giveon et~al.}(1991)\citenamefont{Giveon, Hall, and
  Sarid}}]{Giveon:1991zm}
\bibinfo{author}{\bibfnamefont{A.}~\bibnamefont{Giveon}},
  \bibinfo{author}{\bibfnamefont{L.~J.} \bibnamefont{Hall}}, \bibnamefont{and}
  \bibinfo{author}{\bibfnamefont{U.}~\bibnamefont{Sarid}},
  \bibinfo{journal}{Phys. Lett.} \textbf{\bibinfo{volume}{B271}},
  \bibinfo{pages}{138} (\bibinfo{year}{1991}).

\bibitem[{\citenamefont{Shtabovenko et~al.}(2016)\citenamefont{Shtabovenko,
  Mertig, and Orellana}}]{Shtabovenko:2016sxi}
\bibinfo{author}{\bibfnamefont{V.}~\bibnamefont{Shtabovenko}},
  \bibinfo{author}{\bibfnamefont{R.}~\bibnamefont{Mertig}}, \bibnamefont{and}
  \bibinfo{author}{\bibfnamefont{F.}~\bibnamefont{Orellana}}
  (\bibinfo{year}{2016}), \eprint{1601.01167}.

\bibitem[{\citenamefont{Mertig et~al.}(1991)\citenamefont{Mertig, Bohm, and
  Denner}}]{Mertig:1990an}
\bibinfo{author}{\bibfnamefont{R.}~\bibnamefont{Mertig}},
  \bibinfo{author}{\bibfnamefont{M.}~\bibnamefont{Bohm}}, \bibnamefont{and}
  \bibinfo{author}{\bibfnamefont{A.}~\bibnamefont{Denner}},
  \bibinfo{journal}{Comput. Phys. Commun.} \textbf{\bibinfo{volume}{64}},
  \bibinfo{pages}{345} (\bibinfo{year}{1991}).

\bibitem[{\citenamefont{Hahn and Perez-Victoria}(1999)}]{Hahn:1998yk}
\bibinfo{author}{\bibfnamefont{T.}~\bibnamefont{Hahn}} \bibnamefont{and}
  \bibinfo{author}{\bibfnamefont{M.}~\bibnamefont{Perez-Victoria}},
  \bibinfo{journal}{Comput. Phys. Commun.} \textbf{\bibinfo{volume}{118}},
  \bibinfo{pages}{153} (\bibinfo{year}{1999}), \eprint{hep-ph/9807565}.

\end{thebibliography}
\end{document}